\documentclass[aps,showpacs,preprintnumbers,amsmath,amssymb,superscriptaddress,floatfix,a4paper,nofootinbib,11pt,floatfix,nofootinbib,onecolumn]{revtex4}
\usepackage{graphicx}
\usepackage{bm}
\usepackage{rotating}
\usepackage{array}
\usepackage{xcolor}
\usepackage{amsmath,amssymb}
\usepackage{mathrsfs}
\usepackage{graphicx}
\usepackage{color}
\usepackage{subfigure}
\usepackage{fancyhdr}
\usepackage{multirow}
\usepackage{float}
\usepackage{epsfig}
\usepackage{amsfonts}
\usepackage{bm}
\usepackage{multirow}
\usepackage{nicefrac}
\newcommand{\ba}{\begin{eqnarray}}
\newcommand{\ea}{\end{eqnarray}}

\usepackage{bbm}
\newcommand{\be}{\begin{eqnarray}}
\newcommand{\ee}{\end{eqnarray}}
\usepackage{slashed}

\begin{document}

\title{AdS to dS phase transition mediated by thermalon in Einstein-Gauss-Bonnet gravity from R\'{e}nyi statistics}

\author{Daris Samart} 
\email{darisa@kku.ac.th}
\affiliation{Khon Kaen Particle Physics and Cosmology Theory Group (KKPaCT), Department of Physics, Faculty of Science, Khon Kaen University, 123 Mitraphap road, Khon Kaen, 40002, Thailand}

\author{Phongpichit Channuie} 
\email{channuie@gmail.com (Corresponding~author)}
\affiliation{College of Graduate Studies, Walailak University, Thasala, \\Nakhon Si Thammarat, 80160, Thailand}
\affiliation{School of Science, Walailak University, Thasala, \\Nakhon Si Thammarat, 80160, Thailand}

\date{\today}

\begin{abstract}

In this work, we present the possible existence of a thermalon phase transition between anti-de Sitter (AdS) to de Sitter (dS) vacua in Einstein-Gauss-Bonnet gravity by considering the R\'{e}nyi statistics. A thermalon changes the asymptotic structure of spacetimes via the bubble nucleation of spherical thin-shells which host a black hole in the interior. All relevant thermodynamical quantities are computed in terms of the R\'{e}nyi statistics in order to demonstrate the possible existence of the AdS to dS phase transition. In addition, we also comment on the behaviors of the phase transitions in the R\'{e}nyi statistics.

\end{abstract}


\maketitle


\section{Introduction}

An explanation of a positive value of the cosmological constant is one of the greatest challenges in physics at the present time. This leads to a study of the phase transition in gravitational physics. Indeed, phase transitions between two competing vacuum energies in a given theory are quite well-known in physical phenomena. These happen when the free energy of the actual vacuum becomes greater than the other owing to a variation of the order parameter of the system. The phase transitions between the two competing vacua have been so far studied in terms of gravitational instantons \cite{Coleman1977,Coleman1980} in which the false vacuum locally decays (tunneling) via the bubble nucleation in the presence of gravity. In addition, the Hawking-Page transition (HP) is a well-known example of the gravitational phase transitions \cite{Hawking:1982dh}. This is the phase transition between thermal AdS space and the AdS black hole. The AdS black hole prefers to be stable if its temperature is higher than the critical temperature. In contrast, the system will be dominated by the thermal AdS when the temperature is lower than the critical temperature. A study of black hole phase transitions provides rich phenomena that lead to a better understanding of the laws of black hole thermodynamics and some useful information on the quantum gravity theory. Moreover, based on the AdS/CFT correspondence paradigm \cite{Maldacena:1997re}, the HP phase transition of the five-dimensional black hole in the AdS spacetime is shown that this phenomenon is related to the confinement/deconfinement phases in a QCD theory \cite{Witten:1998zw,Nojiri:2001aj}. Therefore, a study of the phase transitions of higher-dimensional AdS black holes becomes a hot research topic in high energy physics and it might provide more details and a deeper understanding of the AdS/CFT correspondence. 

More importantly, the coincident existence of different AdS/dS vacua is an attractive feature in several gravitational theories. For example, the scalar fields \cite{Linde:1974at,Veltman:1974au} or $p$-form gauge fields \cite{Aurilia:1980xj,Duff:1980qv} coupled to the curvature theories might provide positive and non-zero vacuum expectation values of the corresponding fields then contribute to the cosmological constant. In addition, there are many other mechanisms that would produce phase transitions between distinct vacua, for instance, a quantum tunneling process via an instanton \cite{Brown:1987dd,Brown:1988kg}, the thermally activated phase transition \cite{Linde:1977mm,Linde:1980tt,Linde:1981zj}. Moreover, a study of phase transition in the AdS and dS black hole thermodynamics has been investigated in several aspects and various models of the higher-order gravity \cite{Nojiri:2017kex,Charmousis:2008kc,EslamPanah:2017yoc,Wei:2012ui,Chakraborty:2015taq,Cvetic:2003zy,Garraffo:2008hu}.

In addition, a so-called ``thermalon" $-$the instanton with finite temperature describing the bubble nucleation that causes the false decay is an interesting mechanism that can thermally stimulate the phase transition through the Euclidean sector of the bubble thin-shell mediation with inclusions of the matter fields in Einstein gravity \cite{Gomberoff:2003zh,Kim:2007ix,Gupt:2013poa}. The thermalon and the relevant calculations were proposed in Ref.\cite{Gomberoff:2003zh}. This bubble-thin shell locates between two regions of the spacetime described by false vacuum decay which hosts the black hole in the interior. On the other hand, the thermalon is considered a thermodynamic phase and described as an intermediate state. At the critical temperature with a finite time, when the thermalon (bubble) forms, it is dynamically unstable and then expands to fill up a whole space at the boundary. Hence, this effectively reduces the vacuum expectation value of the metastable state to the true vacuum. The thermalon was first used to study the false vacuum decay to the true vacuum state hosting a black hole inside. As a result, the cosmological constant of the false vacuum is relaxed to the true vacuum corresponding to the observed value of the cosmological constant. In other words, this reveals the gravitational phase transition from dS to dS spacetime. The materialization of the bubble is caused by the exotic matter in Einstein's theory of gravity, a.k.a., the matter with negative pressure. However, all experiments do not confirm the existence of exotic matter yet. Therefore, it is interesting to consider other sources instead of exotic matter. Interestingly, the higher-order theories of gravity with the vacuum solution are used to study the thermalon mediating the phase transition \cite{Camanho:2012da,Camanho:2015zqa,Camanho:2013uda,Hennigar:2015mco,Camanho:2015ysa,Sierra-Garcia:2017rni,Nojiri:2001pm,Cvetic:2001bk} in various aspects. The results show that the higher-order gravity can thermally active the AdS to dS phase transitions with a vacuum solution whereas the thermalon in Einstein gravity needs the matter fields to proceed with the phase transitions \cite{Gomberoff:2003zh}. Furthermore, the gauge fields are included for studying the phase transitions \cite{Samart:2020qya,Samart:2020mnn}. The results reveal that the profile of the phase transitions does not change and the critical temperature and the Gauss-Bonnet coupling of the phase transitions are decreased by including more types of charges. Since the string theory naturally generates higher-order gravity, a study of the phase transitions might expose interesting phenomena and their consequences in the string theory at low energy regimes. In particular, one may expect to gain a better understanding of the phase transition in the AdS/CFT correspondence. It is well known that the dS/CFT correspondence is less studied and poorer understanding than the AdS/CFT counterpart. Therefore, the AdS to dS phase transitions might be useful to explore more details and clearer pictures of the dS/CFT. Moreover, it is worth to mention about the difference between the standard gravitational instanton (Coleman-De Luccia) and the thermalon mediated phase transition that induced the false vacuum decay. On the one hand, the Coleman-De Luccia instanton induces the phase transition between false (exterior spacetime) and true (interior spacetime) vacuum states by the bubble nucleation. On the other hand, the thermalon bubble nucleation thermally activates the phase transition by changing the false vacuum of the exterior  spacetime to the true vacuum of the interior spacetime which hosts a black hole inside. Let us clarify the fate of the bubble of the true vacuum after the formation. The exterior spacetime can be either dS or AdS. In the dS exterior spacetime scenario, there is an existence of the cosmological horizon. After the bubble of the interior spacetime is popping up, it will acceleratedly expand and finally reaches the cosmological of the exterior dS spacetime filling a whole space with the new phase. Even though, for the second case, the exterior AdS spacetime has no cosmological horizon, i.e., infinitely large, the phase transition would be possible. In this case, the bubble (thermalon) can run to the boundary at infinity with the expansion rate as the speed of light asymptotically. It has been demonstrated in Ref.\cite{Camanho:2013uda} (see section 4.C.2) that the bubble can reach the timelike boundary of the AdS space (exterior spacetime) with a finite time. This is interpreted as the change of the boundary conditions of the AdS to dS. In other words, there is a transition between the AdS solution to the dS one.

In recent years, R\'{e}nyi statistics are greatly received attention for application in various fields of physics, for example, condensed matter physics \cite{Varga:2002,Chen:2012}. In particular, the R\'{e}nyi statistics \cite{Renyi-stat} is more general than the Gibbs-Boltzmann statistics that are usually used in a study microscopy of the equilibrium thermal system and provides a well defined of the non-extensive systems in thermodynamics where the entropy in the R\'{e}nyi statistics is represented in terms of the logarithmic form of the standard (Tsallis) entropy with a non-extensive parameter, $\alpha$ \cite{Renyi-entropy}. According to Hawking's black hole area theorem, Bekenstein proposed that the black hole event horizon and its surface gravity correspond to the entropy and temperature of the black holes (BHs). These lead to formulating the black hole thermodynamics and it is very successful to study the thermodynamics of the black hole in related properties and applications from many points of view. However, the entropy of the black holes is the non-extensive quantities. Therefore, the black hole entropy functional form might be written in terms of the R\'{e}nyi statistics with the non-extensive parameter due to the non-extensive nature of the black holes. Moreover, the R\'{e}nyi statistics is also compatible with the zeroth law of thermodynamics where one can define the empirical temperature properly. Consequently, the R\'{e}nyi statistics have an important role and interesting features in the study of black hole thermodynamics. For instance, the R\'{e}nyi entropy suggests that the Schwarzschild black hole probably gains the positive heat capacity \cite{Czinner:2015eyk} whereas the Schwarzschild black hole always has the negative heat capacity in the standard statistics. This means that the Schwarzchild black hole is stable in the R\'{e}nyi framework. In addition, it has been shown that several types of black holes become thermodynamic stable in terms of the R\'{e}nyi entropy approach \cite{Czinner:2017tjq,Tannukij:2020njz,Promsiri:2021hhv}. As mentioned above, the R\'{e}nyi statistics have several attractive advantages in the study of the thermodynamic system of black holes. This motivates several studies of black hole thermodynamics by using R\'{e}nyi statistics in various aspects \cite{Tannukij:2020njz,Promsiri:2020jga,Bialas:2008fa,Dong:2016fnf,Wen:2016itq,Ghodsi:2018vhq,Ren:2020djc,Nakarachinda:2021jxd,Promsiri:2021hhv,Sriling:2021lpr,Chunaksorn:2022whl,Nakarachinda:2022gsb}. We, therefore, employ the R\'{e}nyi to apply to the thermodynamics of black hole hosted by the true vacuum in this work. 

This work aims to study the possible existence of gravitational phase transition from AdS to dS geometries in the higher order gravity with Einstein-Gauss-Bonnet (EGB) term by using the R\'{e}nyi statistics. In addition, we also compare and comment on the similarities and differences of the thermalon phase transitions between R\'{e}nyi and standard statistics. Let's note here how this phase transition can be possibly addressed using other interesting descriptions, e.g., entanglement entropy, of sub-region in the same geometry instead of the BH entropy. Both BH entropy and Renyi entropies are non-additive quantities and do not satisfy subadditivity equality. In addition, there is no simple derivation of both entropies from a microcanonical point of view by simply counting statistical microstates. Indeed, in many examples, there are no such microcanonical states. However, Ryu-Takayanagi (RT) showed in their paper \cite{Ryu:2006bv} that there is a simple formula to derive the leading term in BH entropy via entanglement entropy using the AdS/CFT approach \cite{Hawking:2000da}. By considering a codimension two hypersurface of the corresponding AdS geometry for any given CFT dual system, RT showed that the entanglement entropy is given by $S_{EE}={Area(\gamma)}/{4G}$, where the $Area(\gamma)$ is the minimal area for such a codimension two surface, and $G$ is the gravitational constant. The phase transitions of the $1^{\rm st}$ and $2^{\rm nd}$ types are studied via this RT entropy in different models, e.g., \cite{Ryu:2006bv,Hawking:2000da,Nishioka:2009un}. A discontinuity in the gradient of $S_{EE}$ via the holographic principle (called holographic entanglement entropy or HEE) proposed by RT coincides with other approaches to phase transitions, including geometrothermodynamics \cite{Bravetti:2012hd}, and fidelity susceptibility \cite{Momeni:2016qfv}. In systems undergoing the second-order phase transitions, such as in high-temperature superconductors, such phase transitions are well-studied via HEE. Additionally, Susskind \cite{Brown:2015bva}, Alishahiha \cite{Alishahiha:2015rta}, and Takayanagi \cite{Rangamani:2016dms} later proposed a codimension one volume as a holographic dual to computational complexity, and this novel holographic quantity proposed a simple mechanism for studying phase transitions in different systems \cite{Momeni:2015iea}.

The content of the paper is organized as follows. In section \ref{s2}, we recall the basic formalism of thermalon dynamics in EGB gravity. A section \ref{s3} is a study of the thermalon phase transition and the relevant thermodynamic quantities of the EGB gravity in terms of the R\'{e}nyi statistics where we will investigate how the free energy and temperature are modified by the R\'{e}nyi thermodynamics and compare the thermalon mediated phase transitions with the standard thermodynamics. We summarize and conclude the results in the last section.

\section{Formalism}
\label{s2}
\subsection{The Einstein-Gauss-Bonnet gravity}
We start this section by recalling the action of the EGB gravity in the vacuum at $d=5$ \cite{Camanho:2012da,Camanho:2013uda,Charmousis:2008kc,Garraffo:2008hu}. The total gravitational action of the EGB theory with its boundary term is read,
\begin{eqnarray}
\mathcal{I} &=& \int_{\mathbb{M}} d^5x\sqrt{-g}\left[ -\,\varepsilon_\Lambda\,\frac{12}{L^2} + R + \frac{\lambda\,L^2}{2}\Big( R^2 - 4\,R_{ab}\,R^{ab} + R_{abcd}\,R^{abcd} \Big) \right] 
\nonumber\\
&& -\, \int_{\mathbb{\partial M}} d^{4}x\sqrt{-h}\left[ K +\lambda\,L^2
\left\{ J - 2\left( \mathcal{R}^{AB} -\frac12\,h^{AB}\,\mathcal{R}\right)K_{AB}\right\}\right] \,,
\label{EGBM-action}
\end{eqnarray}
where $J\equiv h^{AB}\,J_{AB}$ is the trace of the $J_{AB}$ tensor which is composed of the extrinsic curvature, $K_{AB}$ as \cite{Davis:2002gn}
\begin{eqnarray}
J_{AB} = \frac13\left( 2\,K\,K_{AC}\,K_B^C + K_{CD}\,K^{CD}\,K_{AB} - 2\,K_{AC}\,K^{CD}\,K_{DB} - K^2\,K_{AB}\right) ,
\end{eqnarray}
and $\mathcal{R}_{AB}$ is the Ricci intrinsic curvature tensor of the hypersurface, $\Sigma$\,. The spacetime indices of the bulk ($d=5$) and hypersurface ($d=4$) are represented by small and capital Latin alphabets, respectively e.g., $a,\,b,\,c,\,\cdots = 0,\,1,\,2,\,3,\,5$ and $A,\,B,\,C,\,\cdots = 0,\,1,\,2,\,3$\,. In addition, we identify the bare cosmological constant $\Lambda$ as
\begin{eqnarray}
\Lambda = \varepsilon_\Lambda\,\frac{6}{L^2}\,,
\end{eqnarray}
where $\varepsilon_\Lambda = \pm\,1$\, is the sign of the bare cosmological constant and we use the $\varepsilon_\Lambda = +\,1$ (de-Sitter) of the bare cosmological constant in this work. We also employed the normalization of the gravitational constant as $16\pi G_N(d-3)!=1$ Refs.\cite{Camanho:2015zqa,Camanho:2013uda,Hennigar:2015mco}.

Next, we collect the spherically symmetric solution of the EGB gravity and the line element of is given by,
\begin{eqnarray}
ds^2 = -\,f(r)\,dt^2 + \frac{dr^2}{f(r)} + r^2\,d\Omega^2_{3}\,,
\label{line-element}
\end{eqnarray}
where $d\Omega^2_{3}=d\theta^2 + \sin^2\theta\, d\chi^2 + \sin^2\theta\,\sin^2\chi\,d\phi^2$ is the line element of the $3$-dimensional surface. Then the solution of the EGB gravity is written in the simple form as
\begin{eqnarray}
\Upsilon[g] &=& -\frac{1}{L^2} + g + \lambda\,L^2\,g^2 = \frac{\mathcal{M}}{r^{4}}\,.
\label{polynomial-sol}
\end{eqnarray}
Here the relation of the $g(r)$ function to the metric tensor $f(r)$ in Eq.(\ref{line-element}) can be written via the following equation,
\begin{eqnarray}
g &\equiv& g(r) = \frac{1 - f(r)}{r^2}\,.
\label{g-f-relate}
\end{eqnarray}
The parameters $\mathcal{M}$ is related to the black hole ADM mass ($M$) as $
\mathcal{M} = M/8\,\pi$. We refer all the detail derivation of the $\Upsilon$ solution in Refs. \cite{Charmousis:2008kc,Garraffo:2008hu}.
By using the polynomial in Eq.(\ref{polynomial-sol}), one obtains the solutions of $g(r)$ and it is given by
\begin{eqnarray}
g_\pm \equiv g_\pm(r) = -\,\frac{1}{2\,\lambda\,L^2}\left( 1 \pm \sqrt{1+ 4\,\lambda\left[ 1
+ L^2\,\frac{\mathcal{M}_\pm}{r^{4}}\right]}\,\right).
\label{g-pm}
\end{eqnarray}
We will see in the latter that one might identify the branches solutions of the line elements in Eq. (\ref{g-f-relate}) as inner and outer manifolds and are given by 
\begin{eqnarray}
f_\pm \equiv f_\pm(r) = 1 + \frac{r^2}{2\,\lambda\,L^2}\left( 1 \pm \sqrt{1+ 4\,\lambda\left[ 1
+ L^2\,\frac{\mathcal{M}_\pm}{r^{4}}\right]}\,\right).
\label{metric-f-pm}
\end{eqnarray}
In addition, it is worth to note that the effective cosmological constants of two branches of spherical symmetric solutions of the EGB gravity are obtained by setting, $\mathcal{M} = 0$ and they are,
\begin{eqnarray}
f_\pm(r) &=& 1 - \Lambda_\pm^{\rm eff}\,r^2\,,
\nonumber\\
\Lambda_\pm^{\rm eff} &=& -\left(\frac{1 \pm \sqrt{1+ 4\,\lambda}}{2\,\lambda\,L^2}\,\right).
\label{eff-CC}
\end{eqnarray}
More importantly, it is found that only the minus branch $f_-(r)$ allows the black hole solution and the Einstein gravity for $\lambda\to\infty$ is recovered whereas the plus branch $f_+(r)$ encounters the Boulware-Deser (BD) ghost instability due to Eq.(\ref{eff-CC}) and the effective cosmological constant of the $f_+(r)$ diverges for $\lambda\to\infty$ limit. Moreover, the effective cosmological constants in Eq.(\ref{eff-CC}) provide the negative and positive values, respectively. This means the $f_+(r)$ and $f_-(r)$ solutions correspond to AdS and dS spaces. In the latter, We will recognize $f_+(r)$ and $f_-(r)$ branches as the outer and inner manifold when a study of the gravitational phase transition between two solutions is taken into account in the next section.

\subsection{Thermalon dynamics and its stability}
The main purpose of this section is to recap the main properties of the dynamics of unstable bubble thin shell (thermalon) in EGB gravity. This leads to the AdS to dS gravitational phase transition. To investigate the phase transition between AdS to dS spacetimes, we, therefore, divide the manifold of the spacetime into two regions.  We also consider a timelike surface of the manifold in this work. The total manifold is decomposed by $\mathbb{M} = \mathbb{M}_{-} \cup (\Sigma\times\xi) \cup \mathbb{M}_+$. Here $\Sigma$ is the junction hypersurface and is used to connect two regions of the spacetime where $\xi \in [0,1]$ is  the interpolating parameter that connects both regions. The manifolds $\mathbb{M}_+$ and $\mathbb{M}_-$ are outer and inner regions of the manifolds, respectively as well as the metric tensor, $f_\pm(r)$, in  Eq.(\ref{metric-f-pm}) are also used for describing the geometries of the outer and inner manifolds, respectively. Two different line elements of the spacetimes of the AdS outer ($+$) and dS inner ($-$) spacetime are given by,
\begin{eqnarray}
ds_\pm^2 = -f_\pm(r_\pm)\,dt_\pm^2 + \frac{dr_\pm^2}{f_\pm(r_\pm)} + r^2_\pm\,d\Omega_{3}^2\,.
\label{out-in-line}
\end{eqnarray}
Next, we will construct a manifold $\mathbb{M}$ by matching $\mathbb{M}_\pm$ at their boundaries. The boundary of the hypersurfaces $\partial \mathbb{M}_\pm$ is written by
\begin{eqnarray}
\partial \mathbb{M}_\pm := \Big\{ r_\pm = a \,\big |\, f_\pm > 0\,\Big\}\,.
\end{eqnarray}
We parameterize the coordinates of both manifolds as,
\begin{eqnarray}
r_\pm = a(\tau)\,,\qquad\qquad t_\pm = \widetilde{t}_\pm(\tau)\,,
\end{eqnarray}
where $\tau$ is a comoving time parameter of the hypersurface, $\Sigma$, which is used to connect both sides of two manifolds, $\mathbb{M}_\pm$, at the boundaries. Therefore we find
\begin{eqnarray}
ds_\Sigma^2 = -d\tau^2 + a^2(\tau)\,d\Omega_{3}^2\,.
\label{surface-line}
\end{eqnarray}
Having used the above conditions, one obtains the following constraint of the hypersurface line element as
\begin{eqnarray}
1 = f_\pm(a)\left(\frac{\partial\, \widetilde{t}_\pm}{\partial \tau}\right)^2 - \frac{1}{f_\pm(a)}\left(\frac{\partial a}{\partial \tau}\right)^2\,.
\end{eqnarray}
We note that a study of the dynamics of a thermalon and its stabilities have been done in detail in Refs.\cite{Camanho:2013uda,Camanho:2015ysa,Hennigar:2015mco} for the vacuum case and in Refs.\cite{Samart:2020qya,Samart:2020mnn} for the inclusions of the gauge fields. In the following, we will provide relevant ingredients useful for a further study of the thermalon phase transition in the R\'{e}nyi thermodynamics.
\\
Based on Refs \cite{Camanho:2013uda,Camanho:2015ysa}, the continuity of the junction condition across the hypersurface of the vacuum case can be represented in terms of the canonical momenta, $\pi_{AB}^\pm$ as
\begin{eqnarray}
\pi_{AB}^+ - \pi_{AB}^- = 0\,.
\label{junction-eqn}
\end{eqnarray}
In addition, the canonical momentum, $\pi_{AB}$, is obtained by varying the boundary with respect to the induced metric, $h_{ab}$, on the hypersurface, $\Sigma$, i.e. \cite{Davis:2002gn,Thibeault:2005ha},
\begin{eqnarray}
\delta \mathcal{I}_{\partial\mathbb{M}} = -\int_{\partial\mathbb{M}}d^{4}x\,\pi_{AB}\,\delta h^{AB}\,.
\end{eqnarray}
It has been demonstrated in Refs.\cite{Camanho:2013uda,Camanho:2015ysa,Hennigar:2015mco} that the diagonal components of, $\pi_{ab}^\pm$, can be written in terms of the relation between time and spatial parts via the following equation,
\begin{eqnarray}
\frac{d}{d\,\tau}\left( a^3\,\pi_{\tau\tau}^\pm\right) = 3\,a^2\,\dot a\,\pi_{\varphi_i\varphi_i}^\pm\,,\qquad \varphi_i = \varphi_1\,,\,\varphi_2\,,\,\varphi_3 = \theta\,,\,\chi\,,\,\phi\,.
\end{eqnarray}
In addition, a co-moving time component of the canonical momentum, $\pi_{\tau\tau}^\pm$, is written in the compact form as \cite{Camanho:2015zqa,Camanho:2013uda,Camanho:2015ysa},
\begin{eqnarray}
\Pi^{\pm} = \pi_{\tau\tau}^\pm 
&=& \int_{\sqrt{H-g_-}}^{\sqrt{H-g_+}} dx\,\Upsilon'\big[ H- x^2\big]\,,
\label{Pi-time-component}
\end{eqnarray}
where $\Upsilon'[x] = d\Upsilon[x]/dx$ and $H = (1 + \dot a^2)/a^2$\,. We define a variable $\widetilde{\Pi}$ as $\widetilde{\Pi} = \Pi^+ - \Pi^-$\,. Consequently, the continuity conditions of the hypersurface across the boundaries are given by
\begin{eqnarray}
\widetilde{\Pi} = 0 = \frac{d\widetilde{\Pi}}{d\tau}\,.
\end{eqnarray}
In order to study the dynamics of thermalon, a Euclidean sector of the spherical thin-shell, we employ the Wick rotation, i.e. $t\,\to\,i\,t$. This leads to $\dot a^2 \,\to\,-\dot a^2$ and $\ddot a \,\to\,-\ddot a$\,. Taking all assumptions previously, the junction condition in Eq. (\ref{junction-eqn}) can be re-written by,
\begin{eqnarray}
\widetilde{\Pi} = \Pi_+ - \Pi_- = 0\,\quad \Longrightarrow \,\quad \Pi_+^2 = \Pi_-^2\,.
\label{junction-EMGB}
\end{eqnarray}
Implications of the $\Pi_\pm$ in Eq. (\ref{Pi-time-component}) and the metric tensor $f_\pm(a)$ in Eq.(\ref{metric-f-pm}), we obtain,
\begin{eqnarray}
\dot a^2 + \frac{a^{6}}{12\, \lambda\,  L^2 }\,\frac{\left(g_+ \left(2 \,g_+\, \lambda\,  L^2+3\right)^2-g_- \left(2\, g_-\, \lambda \, L^2+3\right)^2\right)}{(\mathcal{M}_+ -\mathcal{M}_-) } + 1 = 0\,.
\end{eqnarray}
According to the result of the above junction condition equation, we represent the continuity equation in terms of kinetic and effective potential energies as
\begin{eqnarray}
\Pi_+^2 = \Pi_-^2\,\quad \Longleftrightarrow \,\quad\, \frac12\,\dot a^2 + V(a) = 0\,,
\label{eff-eom}
\end{eqnarray}
where the effective potential, $V(a)$, of the junction condition equation is given by
\begin{eqnarray}
V(a) &=& \frac{a^{6}}{ 24\, \lambda\,  L^2 \,(\mathcal{M}_+ -\mathcal{M}_-)}
\,\Bigg[ \left(1+ 4\,\lambda\right) g + 4 \left(2 + g \lambda\,  L^2\right)\,\frac{\mathcal{M}}{a^{4}}\Bigg]\Bigg|_-^+ + \frac{1}{2}\,.
\label{V-form2}
\end{eqnarray}
The symbol $\big[ \mathcal{O}\big]\big|_-^+$ is defined by
\begin{eqnarray}
\big[ \mathcal{O}\big]\big|_-^+ \equiv \mathcal{O}_+ - \mathcal{O}_-\,.
\end{eqnarray}
The derivative of the effective potential, $V'(a)$, is directly evaluated and we find,
\begin{eqnarray}
V'(a) &=& \frac{a^5}{4\, \lambda\,  L^2 \left(\mathcal{M}_+ -\mathcal{M}_-\right)}
\,\Bigg[(1+ 4\,\lambda)\,g + 2\,\frac{\mathcal{M}}{a^{4}}\Bigg]\Bigg|_-^+ .
\label{div-eff-potential}
\end{eqnarray}
The effective potential and its derivative are crucial quantities for an analysis of the dynamics and stability of a thermalon. To see the dynamics of the bubble's spherical thin-shell, we just change the effective equation of motion in Eq. (\ref{eff-eom}) to the Lorentzian signature. Moreover, the stability of the thermalon can be checked by applying a Taylor expansion around the thermalon location at $a=a_\star$ in the first order. It has been shown in Refs. \cite{Camanho:2013uda,Camanho:2015ysa,Hennigar:2015mco} that the AdS to dS phase transition can take place with a well-defined range of the parameter in the five-dimensional EGB gravity. All detailed demonstrations and discussions have been studied in the existing literature and we will not repeat them here. In addition, we refer to the detailed derivations of $V(a)$ and $V'(a)$ in Refs. \cite{Camanho:2013uda,Camanho:2015ysa,Hennigar:2015mco,Samart:2020qya,Samart:2020mnn}.

Before closing this section, we solve the thermalon solutions that will be used to study the AdS to dS phase transition in the next section. The solutions of the thermalon configuration at $a=a_\star$ is determined by imposing the conditions, $V(a_\star) = 0 = V'(a_\star)$\,. One gets the solutions of $\mathcal{M}_\pm$ in terms of $g_\pm^\star$, $a_\star$ and $\lambda$, $L$ as,
\begin{eqnarray}
\mathcal{M}_+(g_-^\star,\,a_\star,\,\lambda,\,L^2)\; &\equiv& \mathcal{M}_+^\star
\nonumber\\
&=& \frac{1}{4\, \lambda\,L^2\,a_\star^2}
\,(1 + 4\,\lambda)\,a_\star^{4}\,\big[  \left(3 + 2\,\lambda\,L^2\,g_-^\star \right)a_\star^{2} 
+ 6 \, \lambda\,L^2 \big]\,,
\label{M_+}\\
\mathcal{M}_-(g_+^\star,\,a_\star,\,\lambda,\,L^2)\; &\equiv& \mathcal{M}_-^\star
\nonumber\\
&=& \frac{1}{4\, \lambda\,L^2\,a_\star^2}
\,(1 + 4\,\lambda)\,a_\star^{4}\,\big[  \left(3 + 2\,\lambda\,L^2\,g_+^\star \right)a_\star^{2} 
+ 6 \, \lambda\,L^2 \big]\,,
\label{M_-}
\end{eqnarray}
where $g_\pm^\star \equiv g_\pm (a_\star)$\,.
Then, we will obtain the solution of the $g_\pm^\star = g_\pm(a_\star)$ as functions of $a_\star$, $\lambda$ and $L$ by solving the $\Upsilon[g_\pm] = \mathcal{M}_\pm^\star/a_\star^{4}$ equations. We find
\begin{eqnarray}
g_+^\star &=& -\,\frac{(1 + \mathcal{C}_1) + \sqrt{1 + 4\,\lambda - 2\,\mathcal{C}_1 - 3\,\mathcal{C}_1^2 + 4\,\mathcal{C}_2\,\lambda\, L^2}}{2\,\lambda\,L^2}\,,
\label{g+star}
\\
g_-^\star &=& -\,\frac{(1 + \mathcal{C}_1) - \sqrt{1 + 4\,\lambda - 2\,\mathcal{C}_1 - 3\,\mathcal{C}_1^2 + 4\,\mathcal{C}_2\,\lambda\, L^2}}{2 \lambda\,L^4}\,,
\label{g-star}
\end{eqnarray}
where the coefficients $\mathcal{C}_{1,2}$ are given by
\begin{eqnarray}
\mathcal{C}_1 = \frac{ a_\star^2\, (1+ 4\,\lambda)}{2\, a_\star^2 }\,,
\qquad 
\mathcal{C}_2 = \frac{3\,(1 + 4\,\lambda)\left(a_\star^2 + 2\,\lambda\, L^2\, \sigma \right)}{4\,\lambda\, L^2 \,a_\star^2}\,.
\end{eqnarray}
In the $\lambda\rightarrow 0$ limit, we see that $g_-^\star$ is a finite or stable solution while $g_+^\star$ gives an infinite value or unstable solution. In addition, we need to study the phase transition between asymptotic geometries from AdS (outer, $+$) to dS (inner, $-$) of two manifolds of the spacetime, this means that a condition $g_+^\star \neq g_-^\star$ is needed.

\section{Gravitational phase transition}
\label{s3}
\subsection{Entropy and temperature in the R\'{e}nyi statistics}
We are at the crucial part of the present work. All relevant thermodynamics quantities in a study of gravitational phase transition will be determined in the context of the R\'{e}nyi statistics. First of all, the black hole mass of the inner dS spacetime can be found via,
\begin{eqnarray}
f_-(r_{H}) = 0\,,\;\Rightarrow \; g_-(r_{H}) = \frac{1}{r_{H}^2} \,,
\end{eqnarray}
where $r_H$ is the radius of the existent horizons of spacetime. 
The above equation gives
\begin{eqnarray}
\Upsilon_-\left[ \frac{1}{r_{H}^2}\right] = \frac{\mathcal{M}_-}{r_{H}^{4}}\,,
\end{eqnarray}
and this leads to the dS black hole mass as,
\begin{eqnarray}
\mathcal{M}_- = \lambda\,L^2 + r_H^2 -\frac{r_H^4}{L^2}\,,
\end{eqnarray}
where the bare cosmological constant is related to the dS radius by $\Lambda = 6/L^2$\,. In addition, the exact solutions of the horizon are given by,
\begin{eqnarray}
r_H^4 - L^2\,r_H^2 + L^2\,\big( \mathcal{M}_- - \lambda\,L^2\big) = 0\,,
\label{cubic}
\end{eqnarray}
and we find the black hole event horizon, $r_B$ and the cosmological horizon, $r_C$ as
\begin{eqnarray}
r_B &=& \frac{L}{\sqrt{2}} \left[1-\sqrt{1 + 4\left(\lambda  - \frac{\mathcal{M}_-}{L^2}\,  \right)}\,\right]^{\frac12}\,,
\label{rB}
\\
r_C &=& \frac{L}{\sqrt{2}} \left[1+\sqrt{1 + 4\left(\lambda  - \frac{\mathcal{M}_-}{L^2}\,  \right)}\,\right]^{\frac12}\,,
\end{eqnarray}
with the constraint
\begin{eqnarray}
\mathcal{M}_- < \frac{L^2}{4} + \lambda\,L^2\,.
\end{eqnarray}

Next we consider the entropy $S$ of EGB black hole in the standard statistics and it is given by \cite{Cai:2003kt},
\begin{eqnarray}
S = 4\,\pi \sum_{k=0}^{2}\frac{k\,c_k}{5-2\,k}\left( \frac{1}{r_B^2}\right)^{k-1}\,.
\label{entropy}
\end{eqnarray}
where $c_0=1/L^2$, $c_1=1$ and $c_2 = \lambda\,L^2$. The R\'{e}nyi entropy, $S_R$ is defined from the standard entropy by \cite{Czinner:2015eyk,Tannukij:2020njz,Promsiri:2020jga}
\begin{eqnarray}
S_R = \frac{1}{\alpha}\,\ln\big(1+ \alpha\,S\big)\,,
\label{R-entropy}
\end{eqnarray}
where the $\alpha$ parameter is the non-extensive thermodynamical parameter with $-\infty < \alpha < 1$ and in the limit $\alpha\to 0$, the R\'{e}nyi entropy will reduce to the standard entropy. In addition, the R\'{e}nyi temperature is defined by \cite{Czinner:2015eyk,Tannukij:2020njz,Promsiri:2020jga}
\begin{eqnarray}
T_R^{(-)} = \frac{d\mathcal{M}_-}{d S_R} = \frac{d\mathcal{M}_-}{d r_B}\,\frac{d r_B}{d S_R}\,.
\label{dS-temp}
\end{eqnarray}
We refer to all physically relevant discussions of black hole thermodynamics from the R\'{e}nyi statistics in Refs. \cite{Czinner:2015eyk,Tannukij:2020njz,Promsiri:2020jga,Bialas:2008fa,Dong:2016fnf,Wen:2016itq,Ghodsi:2018vhq,Ren:2020djc}. 
\subsection{AdS to dS phase transition from R\'{e}nyi statistics}
The study of the thermal AdS to dS black hole phase transition has been done in Refs.\cite{Camanho:2015zqa,Camanho:2013uda,Camanho:2012da,Hennigar:2015mco} in the vacuum case with the standard entropy and temperature. Before proceeding with our study in the R\'{e}nyi statistics, we would like to recap a short overview of the thermalon phase transition mechanisms in the literature. We start with two different vacua (AdS and dS) in our scenario. The initial state is thermal anti-de Sitter (AdS) space, whereas the final state is de Sitter (dS) space hosting the black hole inside. The exterior thermal AdS is initially in the false vacuum state when the temperature reaches the critical values then it becomes unstable and decays into a black hole inside the interior dS space (true vacuum) by  thermally activated jumping across the Euclidean potential wall. Here the quasi particle state in the Euclidean sector is called the thermalon. The decay mechanism proceeds through nucleation of the spherical thin-shell bubbles of true vacuum (dS) inside the false vacuum (thermal AdS). This means that when the thermalon pops up at some point of the temperature and continually expands until it fills a whole scenario then the initial asymptotically AdS geometry ends up in the stable dS black hole in a finite time and eventually changing asymptotic geometry to dS space. This leads to the transition from negative to positive cosmological constants. The study of this process in the EGB gravity has been initiated by Refs.\cite{Camanho:2012da,Camanho:2013uda}, and the results have revealed that the thermalon effectively jumped from AdS to dS branch solutions in the EGB gravity with $P \propto e^{-\mathcal{I}_E}$ where $P$ and $\mathcal{I}_E$ being the probability of the decay and the Euclidean action of the difference between initial thermal AdS and the thermalon, respectively. In addition, a reversible process for AdS to dS phase transition cannot be happened see detailed discussions in Refs. \cite{Camanho:2015zqa,Camanho:2013uda,Sierra-Garcia:2017rni,Hennigar:2015mco} as well as a reentrant phase transition process in the black hole thermodynamics \cite{Altamirano:2013ane,Frassino:2014pha} does not occur in this framework.

Ref.\cite{Camanho:2013uda} has proven and analyzed that in the canonical ensemble including the bulk (both inner and outer manifolds) and the surface actions, the Euclidean action of the thermalon configuration ($\mathcal{I}_E$) can be written in terms of the inverse Hawking temperature ($\beta_+$), mass ($\mathcal{M}_+$) of the external observer in the asymptotic thermal AdS and the entropy of the dS black hole (see \cite{Camanho:2015ysa} for a detailed derivation). It takes a simple and compact form as,
\begin{eqnarray}
\mathcal{I}_E = \beta_+\,\mathcal{M}_+ + S\,.
\label{thermalon-action}
\end{eqnarray}
By using the on-shell regularization method by subtracting the thermal AdS space (outer branch solution) contribution as discussed in Refs.\cite{Camanho:2015zqa,Camanho:2013uda,Hennigar:2015mco}, here we adopt the results in Eq. (\ref{thermalon-action}) from the standard statistics to the R\'{e}nyi statistics and then the (Gibbs) free energy, $F_R$ in the canonical ensemble of the thermalon configuration in terms of the R\'{e}nyi thermodynamics is given by,
\begin{eqnarray}
F_R = \mathcal{M}_+ + T_R^{(+)}\,S_R\,.
\label{free-energy}
\end{eqnarray}
In the latter, the free energy of the thermalon will be used to compare to the thermal AdS space where the thermal AdS space is set to zero ($F_R^{\rm AdS} = 0$) since it is used to be the background subtraction \cite{Camanho:2015zqa,Camanho:2013uda,Hennigar:2015mco,Sierra-Garcia:2017rni}.  Moreover, we note that there are former five free parameters in the theory in our study, i.e., $\mathcal{M}_\pm$, $T_R^{(\pm)}$ and $a_\star$\,. But after using relations discussed in \cite{Camanho:2015zqa,Camanho:2013uda,Hennigar:2015mco}, we find that $T_R^{(+)}$ is only one parameter of the thermalon phase transition in the EGB gravity.  

The R\'{e}nyi temperature of the external observer in the asymptotic thermal AdS, $T_R^{(+)}$ can be related to the R\'{e}nyi temperature of the dS black hole in Eq. (\ref{dS-temp}) by
\begin{eqnarray}
T_R^{(+)} = \sqrt{\frac{f_+(a_\star)}{f_-(a_\star)}}\,T_R^{(-)}\,.
\label{T_+}
\end{eqnarray}
Using the black hole event horizon in Eq.(\ref{rB}) and substituting into Eqs.(\ref{M_+},\ref{R-entropy},\ref{T_+}), all of the thermodynamics quantities are written as a function of thermalon radius, $a_\star$ and the thermalon properties and the gravitational phase transitions are ready to study in the thermodynamics phase space.  
\begin{figure}[h]	
	\includegraphics[width=10cm]{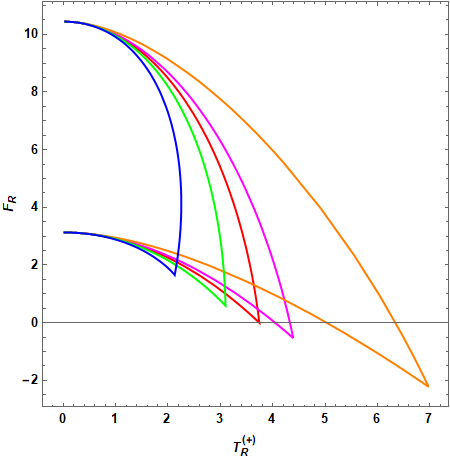}
	\centering
	\caption{Plot displays free energy $F$ of thermalon configuration as function of R\'{e}nyi temperature $T_R^{(+)}$ for several values of non-extensive parameter $\alpha$. We have used $L=1$, and $\lambda =1.1382$. From left to right: $\alpha= -0.25$ (blue), $\alpha =-0.10$ (green), $\alpha=0.00$ (red), $\alpha=0.10$ (magenta) and $\alpha=0.50$ (orange). For each value of non-extensive parameter $\alpha$, the upper branch beyond the cusp is unphysical where it corresponds to $\Pi^+ = -\,\Pi^-$ solutions while the lower branch is physical solutions of $\Pi^+=\Pi^-$. While cusp structures imply the Nariai limit.}
	\label{FT}
\end{figure}

The free energy in Eq.(\ref{free-energy}) has an important role in investigating the phase transition. The profiles of the free energy are characterized by both the coupling $\lambda$ and the non-extensive parameter $\alpha$. We will compare the free energy of the thermalon with respect to the free energy of thermal AdS which is the background subtraction and we set it equal to zero, $F_R^{\rm AdS} = 0$. This means  that when the thermalon free energy in Eq.(\ref{free-energy}) is less than zero the phase transition taking the place. To investigate the AdS to dS phase transition in the R\'{e}nyi statistics, we consider Figure \ref{FT} depicting the plot of free energy $F$ of a thermalon configuration with respect to the R\'{e}nyi temperature $T_R^{(+)}$ for several values of the non-extensive parameter $\alpha$ with a fixed value of the Gauss-Bonnet coupling $\lambda$. It has been shown that the thermalon phase transition in the standard statistics will occur when the coupling $\lambda < \lambda_c = 1.13821$ \cite{Camanho:2015zqa}. In order to see the effects of the R\'{e}nyi thermodynamics on the AdS to dS phase transition, it is very interesting to study a phase transition at $\lambda=\lambda_c$\,. In Figure \ref{FT}, We have fixed $L=1$ and $\lambda=\lambda_c = 1.13821$ from left to right: $\alpha= -0.25$ (blue), $\alpha =-0.10$ (green), $\alpha=0.00$ (red), $\alpha=0.10$ (magenta) and $\alpha=0.50$ (orange). On one hand, in Figure \ref{FT}, the upper branch beyond the cusp corresponds to unphysical branch solutions that stem from $\Pi^+ = -\,\Pi^-$ solutions of the $V(a_\star)=0=V'(a_\star)$ conditions. On the other hand, the lower branch corresponds to the physical solutions from the junction condition, $\Pi^+ = \Pi^-$, see detailed discussions in Ref.\cite{Hennigar:2015mco}. The thermalon temperature and free energy in the physical solutions (lower branch) exist at a certain range of thermalon radius for each value of the Gauss-Bonnet coupling $\lambda$\,. In the case of $\lambda=\lambda_c$, we find $0.51418 < a_\star <0.70711$\,.

We observe that the non-extensive parameter of the R\'{e}nyi entropy, $\alpha$ does modify the phase transition profile. The red plot in figure \ref{FT} corresponds to the free energy at the maximum temperature of the (physical) branch in the standard thermodynamics and the result agrees with the Ref.\cite{Camanho:2015zqa} at $\lambda=1.13821$ and $\alpha=0$ giving $F_R =0$. Increasing positive values of the $\alpha$ make the phase transition at higher R\'{e}nyi temperature (see magenta and orange plots, $F_R < 0$ at maximum temperature) whereas the increasing negative value of the $\alpha$ make the thermal AdS more stable i.e., the phase transition does not take place (see the green and blue plots, $F_R > 0$ at maximum temperature). 
\begin{figure}[h]	
	\includegraphics[width=10cm]{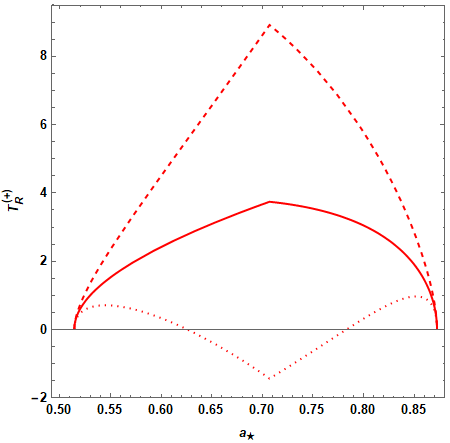}
	\centering
	\caption{Plot shows temperature $T_R^{(+)}$ as function of thermalon radius $a_\star$. Thermalon standard temperature is depicted in solid line while thermalon R\'{e}nyi temperatures are dashed and dotted lines for $\alpha=0.8$ and $\alpha=-0.8$, respectively. Here we have fixed $L=1$, and $\lambda =1.1382$. For each value of non-extensive parameter $\alpha$, the left-handed side of the cusp (the Nariai limit) is physical where it corresponds to $\Pi^+ = \Pi^-$ solutions while the right-handed side beyond cusp is unphysical solutions of $\Pi^+=-\,\Pi^-$. We see that positive $\alpha$ enhances maximum temperature whereas negative $\alpha$ reduces maximum temperature compared to standard statistics.}
	\label{TR}
\end{figure}

In addition, we notice that the range of temperatures over which these phase transitions occur is increasing as the positive magnitude of the non-extensive parameter $\alpha$ is positively increasing with the critical Gauss-Bonnet coupling, $\lambda_c$\,. In contrast, the thermalon mediated phase transitions are not possible when the non-extensive parameter is negative. Moreover, it is observed that when comparing with the impurity substitution, when having charges \cite{Samart:2020qya,Samart:2020mnn}, the present work yields similar results when the non-extensive parameter is positive. We further investigate the range of the non-extensive parameter, $\alpha$ which gives the positive R\'{e}nyi temperature for a range $0.51418 < a_\star <0.70711$ at $\lambda=\lambda_c$\,. Due to the complication of the R\'{e}nyi temperature in Eq.(\ref{T_+}), we therefore perform numerical estimation and we find
\begin{eqnarray}
-0.57886 < \alpha < 1\,.
\end{eqnarray}
The ranges of $\alpha$ given above provide positive values of the R\'{e}nyi temperature. Figure \ref{TR} depicted the thermalon temperature in the standard (solid line) and R\'{e}nyi statistics (dashed and dotted lines for $\alpha =0.8\,,-0.8$ respectively). The results show that increasing non-extensive parameter $\alpha>0$ enlarges the maximum temperature compared to the standard thermodynamics ($\alpha \to 0$); while decreasing non-extensive parameter $\alpha < 0$ reduces the maximum temperature.

\section{Conclusion}

In the present work, we have analyzed the AdS to dS phase transition in EGB gravity with the R\'{e}nyi statistics. A gravitational phase transition of higher-order gravity in a vacuum case has been extensively investigated in the literature using standard thermodynamics. It has been shown that a thermalon, the Euclidean spherical thin-shell, plays an important role in the phase transition mentioning in section \ref{s3}. The thermalon effectively jumps between the branches of the solutions from AdS spacetime to dS spacetime via the thermal activation at a critical value of the Gauss-Bonnet coupling, $\lambda_c$ \cite{Camanho:2015zqa}. This type of phase transition in higher-order gravity is expected as a generic behavior. We, therefore, investigate an extensive analysis of a study of the AdS to dS phase transition by using alternative statistical mechanics to present whether or not the thermalon phase transition in EGB gravity changes its profile. In this work, we employ the R\'{e}nyi statistics as the main framework which has a very interesting feature in a study of black hole thermodynamics. However, the behavior of thermalon dynamics and its stability does not change in the R\'{e}nyi thermodynamics. 

The interesting result in this work is that the non-extensive parameter, $\alpha$ does play the role as the additional order parameter of the thermalon phase transition. The signs of the $\alpha$ parameter affect the phase transition at the critical value of the Gauss-Bonnet coupling as shown in the section \ref{s3}. Interestingly, the consideration of alternative statistics exposes several interesting features of the gravitational phase transitions. In this work, we found that the positive non-extensive parameter, $\alpha>0$ of the R\'{e}nyi thermodynamics enhances the maximum temperature of the thermalon whereas the negative $\alpha<0$ reduces the maximum thermalon temperature compared to the standard statistics in the literature. Since this work is a toy model of the gravitational phase transition in the vacuum of the higher-order gravity, therefore a further study of this work might use for the more physical constraint of the non-extensive, $\alpha$ of the R\'{e}nyi entropy where this parameter is quite difficult to find the physical range of the values. Therefore, an extensive analysis of a more realistic gravitational phase transition will be plausible in studying more detailed properties of the R\'{e}nyi statistics. As the results of this study, our results are compatible with the claim that the gravitational AdS to dS phase transition is a generic transition mechanism of the theories of higher-order gravity with the positive values of the non-extensive parameter, $\alpha$ from the R\'{e}nyi thermodynamics.  

Moreover, an addition of the matter fields is worth investigating to gain a better understanding of the transition between the AdS/CFT correspondence to its dS/CFT counterpart. In particular, some higher spin fields in string theory would yield rich phenomena and new interesting features of the gravitational phase transition. In our present study, the EGB bulk action has a well-defined $4D$ dual boundary theory. Therefore, it is expected that by using HEE approach via the RT formalism, we can observe the same type of phase transitions. The difference is that for the dS path, we need to use a different modified version of HEE, see for example \cite{Hubeny:2007xt}. In summary, we can extend the present work by calculating HEE and thus the implementation can be considered as an alternative method to study the phase transitions.

\acknowledgments
D. Samart is supported by the Fundamental Fund 2565 of Khon Kaen University and DS has received funding support from the National Science, Research and Innovation Fund. P. Channuie acknowledges the Mid-Career Research Grant 2020 from National Research Council of Thailand (NRCT5-RSA63019-03). This work is partially funded by the Thailand Science Research and Innovation (TSRI) under the Program Management Unit (PMUB) with grant No.B05F650021.

\subsection*{Data availability statement}
Data will be made available on reasonable request.

\end{document}